\documentclass[amsmath, amssymb, preprintnumbers,showpacs,showkeys,aps,prb,twocolumn]{revtex4-1} %
\usepackage{graphicx}
\usepackage{float}
\usepackage{color}
\usepackage{braket}
\usepackage{ulem}   % to strike things out
\usepackage[colorlinks=true,citecolor=blue,linkcolor=blue,urlcolor=blue]{hyperref}
\usepackage{mathtools}
\usepackage{bbm}    % for identity matrix

\usepackage{booktabs}
\newcommand{\bs}[1]{{\boldsymbol{#1}}}

\newcommand*{\id}{{\normalfont\hbox{1\kern-0.15em \vrule width .8pt depth-.5pt}}}

\begin{document}
\preprint{}
\title{Influence of point defects on the electronic and topological properties of monolayer WTe$_2$}
% \title{Stability of the 2D topological insulator phase in monolayer WTe$_2$: ab-inito calculation of point defects}
% \title{Effect of defects on the topological properties of the 2D topological insulator WTe$_2$}
\author{Lukas Muechler}
\affiliation{Center for Computational Quantum Physics, The Flatiron Institute, New York, New York, 10010, USA}
\email{lmuechler@flatironinstitute.org}
\author{Wei Hu}
\affiliation{Hefei National Laboratory for Physical Sciences at Microscale,
University of Science and Technology of China, 
Hefei, Anhui 230026, China}
\affiliation{Computational Research Division, Lawrence Berkeley National Laboratory, Berkeley, CA 94720}
\author{Lin Lin}
\affiliation{Department of Mathematics, University of California, Berkeley, CA 94720}			
\affiliation{Computational Research Division, Lawrence Berkeley National Laboratory, Berkeley, CA 94720}
\author{Chao Yang}
\affiliation{Computational Research Division, Lawrence Berkeley National Laboratory, Berkeley, CA 94720}
\author{Roberto Car}
\affiliation{Department of Chemistry, Princeton University, Princeton, New Jersey 08544, USA}
\date{\today}

\begin{abstract}
In some topological insulators, such as graphene and WTe$_2$, band inversion originates from chemical bonding and space group symmetry, in contrast to materials such as Bi$_2$Se$_3$, where the band inversion derives from relativistic effects in the atoms.
In the former, band inversion is susceptible to changes of the chemical environment, e.g. by defects, while the latter are less affected by defects due to the larger energy scale associated with atomic relativistic effects.
Motivated by recent experiments, we study the effect of Te-vacancies and Te-adatoms on the electronic properties of WTe$_2$. We find that the Te-vacancies have a formation energy of $2.21$ eV, while the formation energy of the Te-adatoms is much lower with $0.72$ eV.
The vacancies strongly influence the band structure and we present evidence that band inversion is already reversed at the nominal composition of WTe$_{1.97}$. In contrast, we show that the adatoms do not change the electronic structure in the vicinity of the Fermi level and thus the topological properties. Our findings indicate that Te-adatoms should be present in thin films that are grown in a Te-rich environment, and we suggest that they have been observed in scanning tunneling microscopy experiments.

\end{abstract}
%\pacs{71.20.Lp, 79.60.-i}
\maketitle

\section{Introduction} The advent of topological band theory led to the discovery of a plethora of new topological phases in three-dimensional (3D) materials, facilitated by the extensive databases available from experiment.~\cite{topoquant,ashvin,robert}
In contrast, stable two-dimensional (2D) topological materials are still largely unexplored as the available databases are comparatively much smaller.
Recently, however, monolayers of transition metal dichalcogenides (TMD) of the general formula $MX_2$, such as the \mbox{type-II} Weyl semimetal~\cite{type2wte} WTe$_2$, have been theoretically predicted to be 2D topological insulators (TI).~\cite{LiangFuWTe2,muechler,Ok2019custodial}
Monolayers of TMDs have attracted significant interest in recent years due to their diverse electronic and optical properties that can be used in technological applications.\cite{tmdrev1}
\begin{figure}[t]
\centering,
\includegraphics[width=0.9\columnwidth]{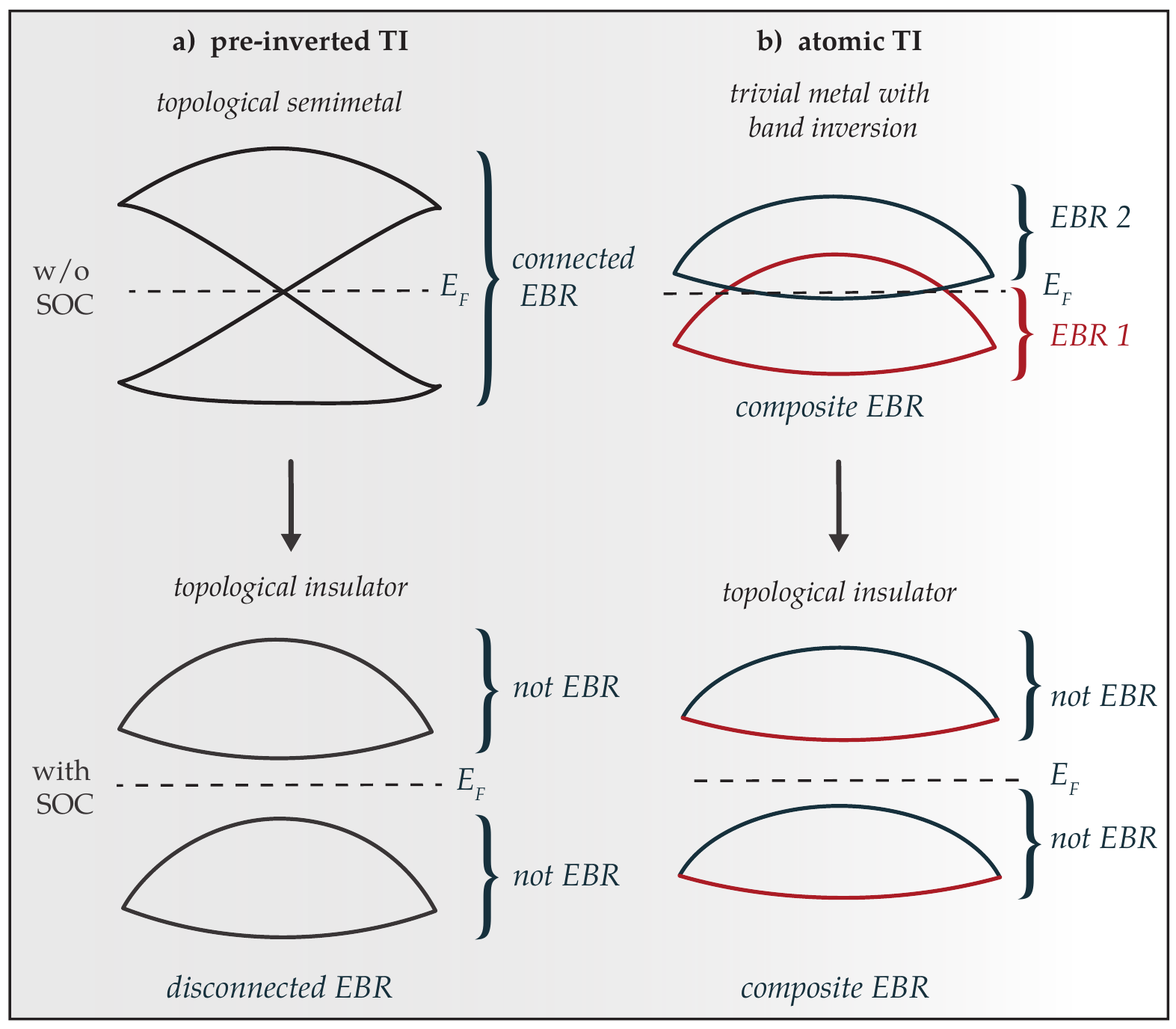}
\caption{
(a) Schematic band structure of an \textit{pre-inverted} TI such as WTe$_2$ or graphene. Without SOC, the material is a topological semimetal with a Dirac crossing , i.e. the band structure forms a connected EBR (see text for explanation). Upon inlusion of SOC, the Dirac cone gaps and the bands split into two disconnected sets. All four bands form a disconnected EBR, while none of the pairs alone does. If only one of the disconnected sets is occupied, the  material is a TI.
(b) Schematic band structure of an \textit{atomic}-TI, such as HgTe or Bi$_2$Se$_3$ that derives from a band inversion between two different EBRs.
% Without SOC the band structure is gapped and forms a composite EBR, rendering the material a topologically trivial insulator.
When scalar relativistic effects such as the Darwin-term are above a threshold, two EBRs can overlap and the band structure is metallic, forming a composite EBR.
Upon consideration of SOC, the band crossings gap out and an insulator is formed. All four bands still form a composite EBR, while none of the pairs is an EBR. Therefore the band structure is topological, if only one set of bands is occupied.}
\label{fig:zero}
\end{figure}
Shortly after the theoretical prediction, many groups have reported fabrication of thin films of WTe$_2$ \cite{synth1,synth2,synth3,wte2leslie,bobpablo} and have already found experimental evidence for topologically protected edge states in monolayer devices.~\cite{qsh1,qsh2,qsh3,qsh_bob}
Monolayers of WTe$_2$ are ideal model materials to study 2D topological phases due to their large band gap. In addition, the interplay of spin-orbit coupling (SOC), low crystal symmetry and high electron mobilities gives rise to other phenomena of interest such as large spin-orbit torques that could be used in spintronic devices.~\cite{SOT1,SOT2} \\

General topological arguments suggest that a TI should not be affected by disorder, as long as the inversion between valence and conduction bands is maintained and a bulk band gap exists, while time-reversal symmetry (TRS) is preserved on average~\cite{RachelMass}. 
This statement is expected to hold for TIs in which the band inversion is due to \textit{atomic} relativistic effects, which are only weakly dependent on the crystal structure and the chemical environment, as is the case with HgTe, Bi$_2$Se$_3$, KHgSb and monolayers of Bi.~\cite{BIsoc,muechler2,Bi2Se3_haijun,wang2016hourglass,ronnyBi} 
In contrast, there is another class of TIs, such as graphene and monolayers of WTe$_2$, where band inversion is not due to relativistic effects, but to the chemical environment and symmetry, which we will call \textit{pre-inverted} TIs.~\cite{QSHgraph,muechler,3dchemdirac}  

One can make this distinction more precise in the language of elementary band representations (EBRs) introduced by~\citet{topoquant}:
A set of energy bands $\{\varepsilon (\bs{k})\}$ is an EBR if it derives from a given collection of localized Wannier functions and cannot be decomposed into a set of other band representations. For example, the set of bands making up the Dirac cone in graphene represents an EBR induced by the carbon $p_z$ orbitals. 
A disconnected set of bands, i.e. a set of bands that is separated from other bands by a gap is topological, if it cannot be derived from an EBR, i.e. if it cannot be derived from a set of localized Wannier functions.
In 'pre-inverted' TIs such as graphene and WTe$_2$, the valence and conduction bands form a connected EBR that decomposes into a disconnected one once SOC, even infinitesimal, is considered [Fig.~\ref{fig:zero}(a)].
While the sum of both valence and conduction bands is an EBR, neither the valence band nor the conduction band alone are an EBR. 
Therefore, if only the lower set of bands is occupied, the system has to be a TI.\\
In contrast, 'atomic TIs' are gapped without SOC, as the valence and conduction band form a composite band representation [Fig.~\ref{fig:zero}(b)].~\cite{schoop2018chemical}
In these compounds, scalar relativistic effects such as the Darwin term are strong and EBRs invert to form a composite, metallic  EBR. Once SOC is considered, a gap opens while the EBR remains composite. However, the disconnected components are topological.~\cite{BiHgTe} \\
In pre-inverted TIs such as WTe$_2$, changes of the local bonding due to defects can reverse the band inversion and drive the material into a topological trivial state.
The energy scale associated to this transition is expected to be small compared to energy scale of atomic relativistic effects.
For example, in the simplest model of a 'pre-inverted' TI, the Kane-Mele model, the energy scale of the phase transition to a topologically trivial insulator is of the order of the next-nearest neighbor hopping term.~\cite{RachelMass} 
For this particular class of materials it is essential to understand the influence of defects on the electronic structure. \cite{Wte2degration,degrationO2,bobpablo} \\

In this paper, we study the effect of Te-vacancies and Te-adatoms on the topological and electronic properties of monolayer WTe$_2$. This is motivated by recent experiments, which have shown that Te-vacancies have drastic effects on the properties of thin WTe$_2$ films, such as a much lower mobility, increased scattering rates and much higher resistance at lower temperature.~\cite{Tedefects} 
In contrast, a recent scanning tunneling microscopy (STM) study showed the existence of topological edges states on step edges of WTe$_2$, despite the presence of unidentified defects.~\cite{stm2}
We propose that these defects are Te-adatoms which possess a low defect formation energy of $0.72 eV$ in Te-rich conditions. These findings are supported by the fact that monolayers of WTe$_2$ are usually obtained from bulk crystals of WTe$_2$ grown in a Te-rich environment~\cite{qsh_bob,teflux}.
Our calculations show that Te-vacancies could drive the system into a trivial state, while the effect of Te-adatoms can be considered as a weak perturbation to the 2D TI state in accordance with experiment.
We show that this distinction occurs because vacancies and adatoms have a different effect on the chemical bonding responsible for band inversion in WTe$_2$.

\section{Monolayer WTe$_2$}
\begin{figure}[t]
\centering
\includegraphics[width=1\linewidth]{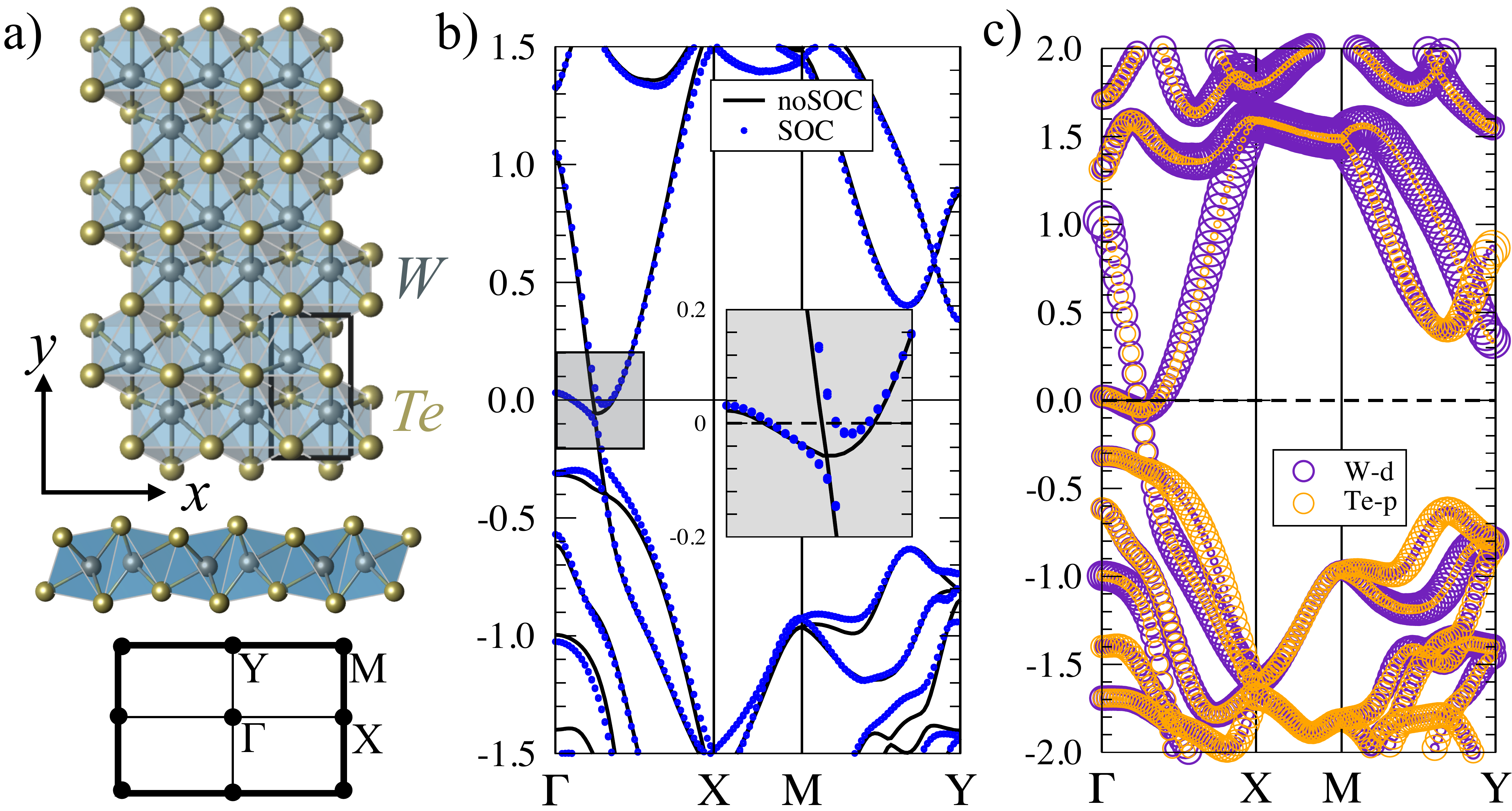}
\caption{
(a) Top and side view of a WTe$_2$ monolayer in the $1T'$ structure and Brillouin zone with high symmetry points.
(b) Band structure of WTe$_2$ monolayers with and without SOC.
(c) Orbital contributions to the band structure of WTe$_2$ around the Fermi energy.
}\label{fig:one}
\end{figure}

% Monolayer TMDs most commonly occur in two structure types, the hexagonal $1H$ structure with $MX_6$ trigonal prisms and the $1T$ structure with edge-sharing $MX_6$ octahedra. Due to the large electronegativity (EN) differences between the $M$ and $X$ atoms, most monolayer TMDs are insulating due to a band gap between the filled $X$-$p$ orbitals and the empty $M$-$d$ orbitals close to the $K$ point. 
TMDs with heavy elements such as WTe$_2$ occur in distorted $1T$ structures. This distortion can be attributed to relativistic effects that determine the relative level alignment of the $p$- and $d-$orbitals.~\cite{hoffmann} In the case of WTe$_2$, the structure distorts from the $1T$ to the orthorhombic $1T'$ structure in which the W atoms form linear chains [Fig.~\ref{fig:one}(a)]. 
In absence of SOC, a monolayer of WTe$_2$ in the $1T'$ structure is a topological Dirac semimetal with two Dirac crossings along $X\Gamma X$,protected by a glide symmetry along the W-chain direction. Similar to graphene, the band structure is already inverted. Upon inclusion of SOC the Dirac cones gap and the system becomes a 2D-TI [Fig.~\ref{fig:one}(b)].~\cite{muechler,LiangFuWTe2,Ok2019custodial} \\
% \red{
% Experimentally, monolayers of WTe$_2$ on substrates such as bilayer graphene are found to be insulating by ARPES and transport experiments. 
% }
In contrast to other TMDs, where the band gap  is between filled $M-d$ orbitals and unoccupied $X-p$ orbitals,
the W-$d$ states are strongly hybridized with the Te-$p$ states around the Fermi energy ($E_F$) in WTe$_2$, due to the small difference in electronegativity between W and Te [Fig.~\ref{fig:one}(c)].
The bonding in WTe$_2$ is therefore quite different from other TMDs such as MoS$_2$. We thus expect WTe$_2$ to be especially susceptible to defects or disorder that disrupts the bonding between W and Te atoms. 

\section{Effect on the electronic structure}
\begin{figure*}[t!]
\centering
\includegraphics[width=0.75\linewidth]{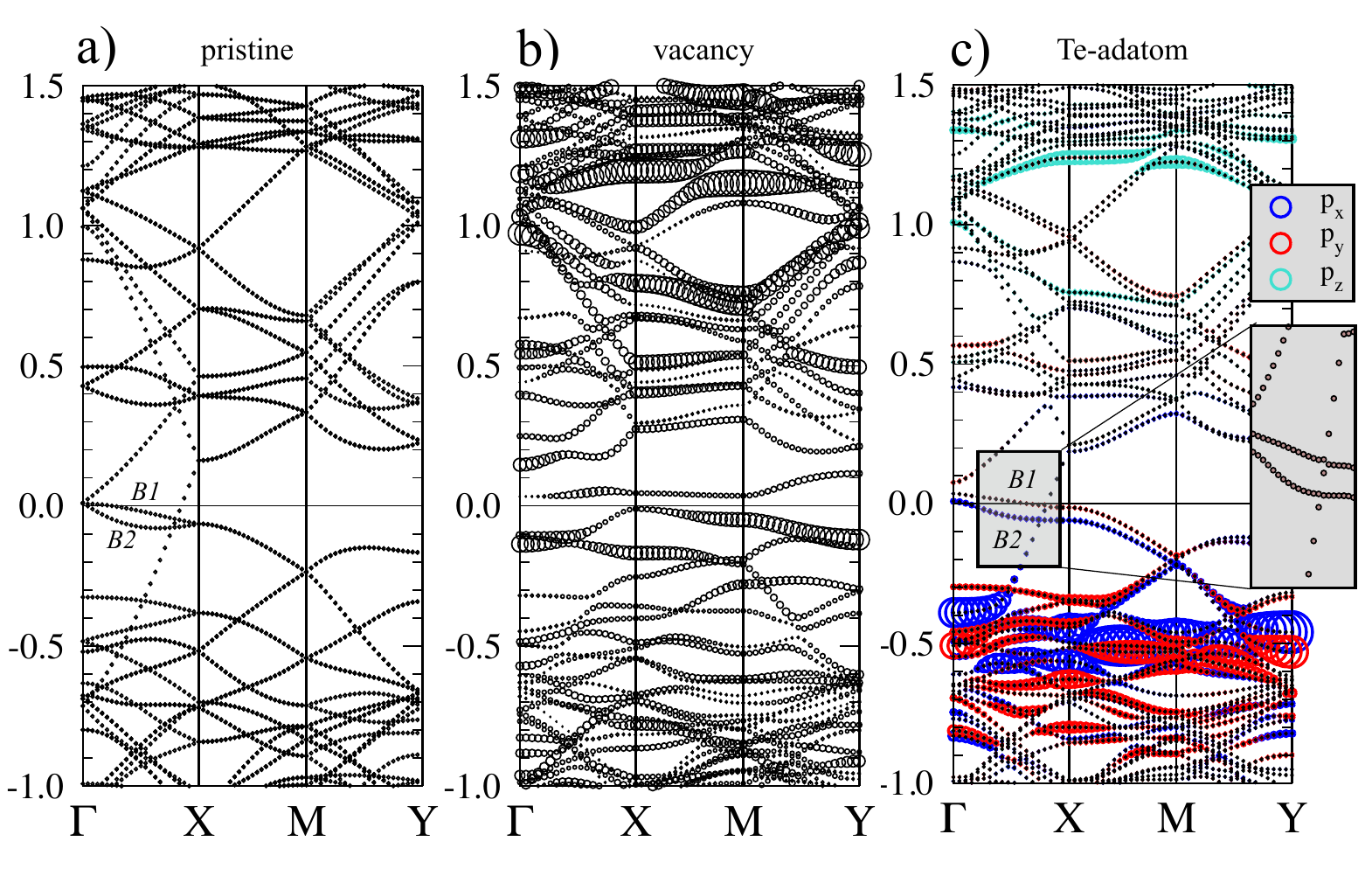}
\caption{
Band structure of a $5\times3$ supercell (no SOC) with 
(a) no defect.
(b) Te-vacancy. The size of the circles indicate the contribution of the W-atoms next to the vacancy.
(c) Te-adatom. The size of the circles indicate the contribution of the adatom states to the band structure. Green circles represent $p_z$, blue $p_x$ and red $p_y$ states. The labels $B1$ and $B2$ are referred to in the text.
}\label{fig:two}
\end{figure*}
We now discuss the influence of two experimentally relevant Te-defects, namely Te-adatoms and Te-vacancies, on the electronic structure and topological properties of WTe$_2$. These defects have been chosen since they are relevant for ongoing experimental work, e.g. Te-vacancies have been shown to influence the transport properties of thin WTe$_2$ films significantly.~\cite{Tedefects}
To understand the influence of these defects on the band inversion, we calculated the band structures without SOC for a fully relaxed $5\times3$ supercell of (i) pristine WTe$_2$, (ii) a Te-vacancy with a formation energy of $E(V) = 2.21 \ eV$ and (iii) a Te-adatom with a formation energy of $E(A) = 0.72 \ eV$.
The defect formation energies were calculated in Te-rich conditions taking into account SOC as \mbox{$E(A) = E'(A) - \left[ E + \mu_{Te} \right]$} for the adatom and
\mbox{$E(V) = E'(V) - \left[ E - \mu_{Te} \right]$} for the vacancy, where $E$ is the energy of the pristine supercell and $E'(A/V)$ is the energy of the supercell with adatom/vacancy, while $\mu_{Te}$ is the chemical potential of crystalline Tellurium (see App.~\ref{app} for computational details).\\
In the pristine supercell, band backfolding leads to the appearance of two Dirac crossings along $\Gamma X$ close to $E_F$, while a large gap separates valence and conduction bands in the rest of the Brillouin zone [Fig.~\ref{fig:two}(a)].
By removing a Te-atom, the band structure changes dramatically [Fig.~\ref{fig:two}(b)]. The Dirac-crossings disappear and a gap is opened along the $\Gamma X$ line with almost flat bands close to $E_F$. By plotting the atomic weights of the W-atoms surrounding the vacancy, we see a large contribution of these atoms to the flat bands close to $E_F$, indicating a localized set of states that does not disperse. This suggests that the W-$d$ states around the vacancies form non-bonding dangling bond states, consistent with the large formation energy of this defect.
The formation of a vacancy disrupts the network of strong covalent, almost metallic bonds. Since the Dirac crossings stem from both W-$d$ and Te-$p$ states, a vacancy cannot considered to be a weak perturbation, as it significantly and qualitatively alters the electronic structure close to $E_F$.
 % In our calculation, we assume a periodic arrangement of defects by using a supercell with a single defect, and only qualitative statements about the influence of the defects on the electronic structure can be made. 
\label{sec:effect}
In contrast, the electronic structure of WTe$_2$ with an Te-adatom is changed only slightly close to $E_F$. [Fig.~\ref{fig:two}(c)]
The interaction of the adatom with the layer leads to a splitting of the atomically degenerate $p$-orbitals. Due to the 2D nature of the monolayer, the in-plane $p_x$ and $p_y$ orbitals of the adatom  interact differently with the monolayer and are lower in energy with respect to the out-of-plane $p_z$ orbital. A neutral Te atom possesses four $p$ electrons and we therefore expect the two $p_x$ and $p_y$ adatom states to be occupied and the $p_z$ states to be unoccupied.
This is reflected in the atomic weights of the adatom states in Fig.~\ref{fig:two}(c), where the $p_x,p_y$ states are clearly separated from the unoccupied $p_z$ states by over 1.5 eV. Due to the large splitting between the in- and out-of-plane orbitals, the adatom states do not contribute to the states close to $E_F$ and the band crossing remains largely unaffected.
The main effect of the adatom is the breaking of the translational symmetries $\{ E | \frac{1}{5} 0 0\}$ and $\{ E | 0 \frac{1}{3} 0\}$, which are the trivial translational symmetries obtained by creating a $5\times3$ supercell.
These symmetries lead to non-symmorphic degeneracies at high-symmetry points due to the back-folding of the bands in the supercell.
Breaking of this symmetry slightly gaps the degeneracies induced by the band folding, but does not significantly change the band crossings close to $E_F$. The strong covalent bonds that contribute to the Dirac cone in WTe$_2$ stem from strong in-plane bonds. An adatom that perturbs the bonding perpendicular to the plane can therefore be treated as a weak perturbation to the band structure, in contrast to the Te-vacancy, which strongly disrupts the in-plane bonding and the states close to $E_F$.
\begin{figure*}[t]
\centering
\includegraphics[width=0.75\linewidth]{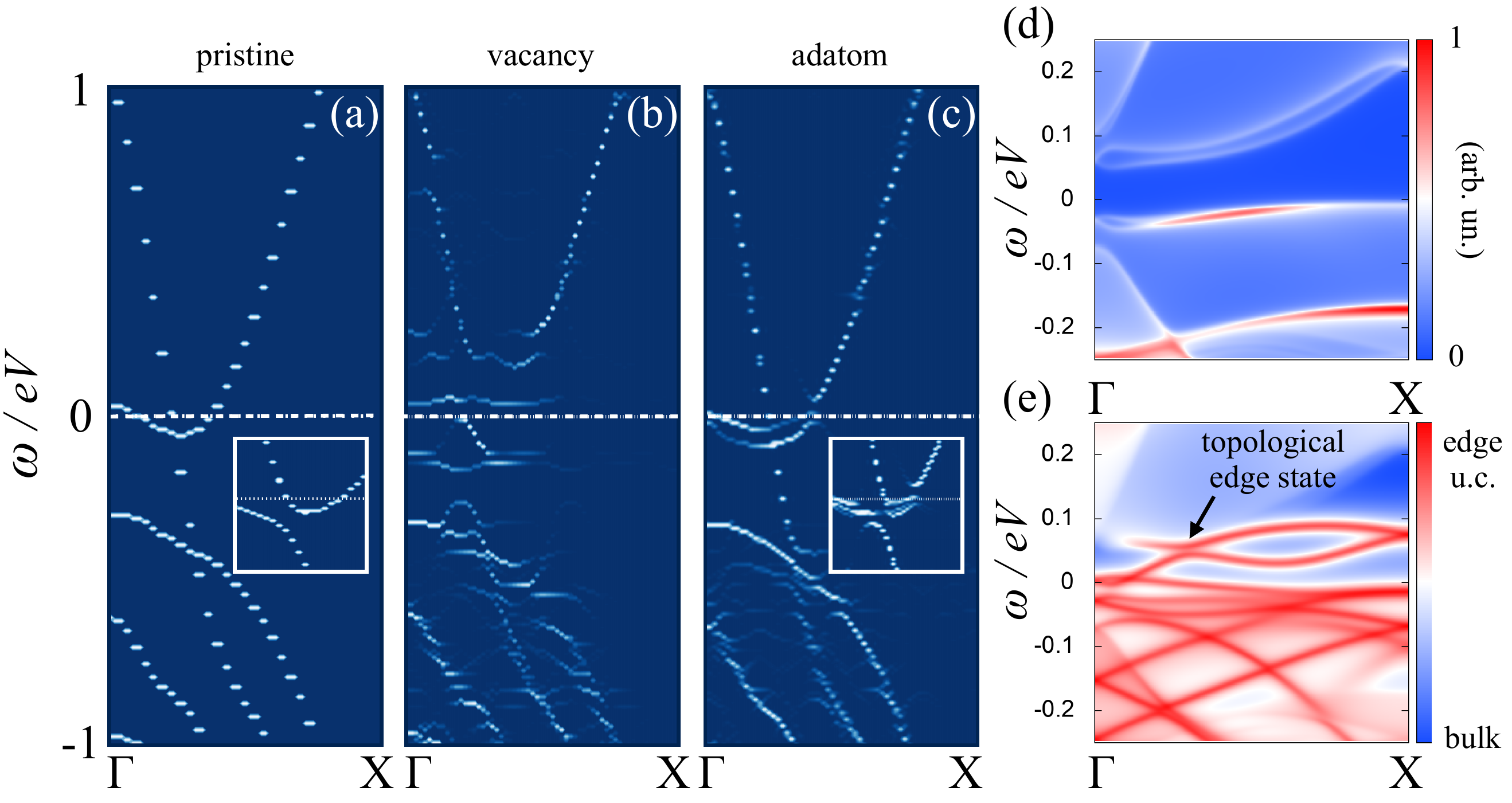}
\caption{
Spectral function of the (a) pristine $5\times3$ supercell, (b) $5\times3$ vacancy supercell  and  (c) $5\times3$ adatom supercell  folded back to the primitive unit cell. The effect of SOC on the states close to the Fermi level are shown in the  insets (The plots have been obtained via the \textit{bandup}-code~\cite{bandup1,bandup2}.)
Spectral functions of a  ribbon constructed from $N = 15$ supercells ($ 4\times2$) with Te-adatoms projected on (d) the bulk states and (e) on the supercell at the edge, calculated from a Wannier-interpolation with SOC.
}\label{fig:three}
\end{figure*}
We confirm these conclusions by calculating the spectral functions of the supercells, which have been back-folded into the primitive BZ along the $\Gamma-X$ direction using the \textit{bandup}-code.~\cite{bandup1,bandup2} The spectral function is given by $\mathcal{A}(\bs{k},\omega) = \sum_{n} P_{n\bs{K}}(\bs{k}) \delta(\omega-\varepsilon_n (\bs{k}))$, where the spectral weight $P_{n\bs{K}}(\bs{k}) = \sum_m |\braket{m\bs{K}|n\bs{k}}|^2$ is defined via the overlap of a state $\ket{m\bs{K}}$ of the supercell with the state $\ket{n\bs{k}}$ at energy $\varepsilon_n (\bs{k})$ of the primitive cell; the k-vectors $\bs{K}$ of the supercell are related to the k-vectors $\bs{k}$ of the primitive cell by a unique folding-vector $\bs{G}$ via $\bs{K} = \bs{k} - \bs{G}$.
The back-folding of the spectral function of the pristine supercell [Fig.~\ref{fig:three}(a)] into the primitive BZ reproduces the well-known band-structure of WTe$_2$.
Introducing a Te-vacancy [Fig.~\ref{fig:three}(b)] strongly perturbs the band structure and leads to a suppression of spectral weight for a large range of energies as shown in Fig.~\ref{fig:two}(b). The spectral function close to $E_F$ is very different from that of the pristine material. The band crossings are absent, indicating that the Te-vacancy drives the system into a topologically trivial state that is not perturbatively connected to the pristine band structure. 

In case of the Te-adatom [Fig.~\ref{fig:three}(c)], the spectral function close to the Fermi-level changes slightly relative to the pristine material due to a small band splitting that generates a pocket near the $\Gamma$-point. However, the Dirac-crossing is left unperturbed and the system should be a 2D TI upon inclusion of SOC. The band splitting is due to the symmetry breaking induced by the adatom. We have highlighted two supercell bands denoted by $B1$ and $B2$ in Fig.~\ref{fig:two}(a) and (c), which fold into the less dispersive band of the Dirac cone in the primitive BZ.
Without adatom, these two bands are degenerate at $X$ and $\Gamma$ and therefore fold back into the same band. In presence of the adatom, the degeneracies are lifted and the bands do not fold back into one unique band in the supercell, leading to a splitting of spectral weight between the two now inequivalent bands.
Upon consideration of SOC [Fig.~\ref{fig:three}(d)], we find that the Dirac cone gaps as expected. The states close to $E_F$ remain broadened due to the symmetry breaking effect of the adatom discussed above. A spectral gap between the valence and conduction band remains in which we expect topological edge states to appear.

\section{Topological properties} 
A monolayer of WTe$_2$ without SOC possesses an inverted band structure with two Dirac cones along $X\Gamma X$, which guarantees the 2D TI phase upon inclusion of SOC due to the presence of the in-plane glide symmetry. \cite{muechler}
Point defects will necessarily destroy the non-symmorphic symmetry and thus a gap opening via SOC does not guarantee a 2D TI state in the presence of defects. Perturbative changes to the band structure without SOC, that leave band-inversions and Dirac cones intact, are not expected to change the topological invariant of WTe$_2$, as long as they do not close the bulk band gap. To confirm that the topological structure is indeed left-intact by the presence of adatoms, we calculated the edge spectral function of a ribbon periodic in the $\hat{x}$-direction, constructed from $N = 15$ supercells ($ 4\times2$) with one Te-adatom per supercell. The spectral function has been calculated from a tight binding model that we obtained from a Wannier interpolation of the DFT band structure with SOC.~\footnote{We chose a $4\times2$ supercell since we could not obtain a Wannier hamiltonian for the $5\times3$ supercell.}
Fig.~\ref{fig:three}(d) shows the bulk projection of the spectral function which displays a well defined bulk band gap. As expected, a surface state connecting valence and conduction bands can be observed clearly in Fig.~\ref{fig:three}(e), where the spectral function has been projected on the edge, proving that the adatoms do not change the topological nature of the slab.
Strong perturbations that gap the Dirac cones on the other hand may lead to a transition to a topologically trivial state.
% We find, that the perturbations with negative formation energies such as Te-adatoms will not change the 2D TI state of monolayer WTe$_2$. 
Our calculation shows that a vacancy concentration of about one percent could already be sufficient to destroy the topological state. \\ 
Furthermore, it is expected that the adatom defects will not serve as strong scattering centers, since they leave the states close to $E_F$ almost unchanged, and there is no contribution from adatom orbitals. Vacancies on the other hand are expected to scatter the conduction electrons strongly. 
Our findings suggest that a systematic study of electron mobilities as a function of adatom and vacancy defect concentration could be of great interest in thism material, which can be induced by changing the growth conditions \cite{Tedefects} or by other techniques such as ion bombardment. 

\begin{figure}[t]
\centering
\includegraphics[width=0.8\linewidth]{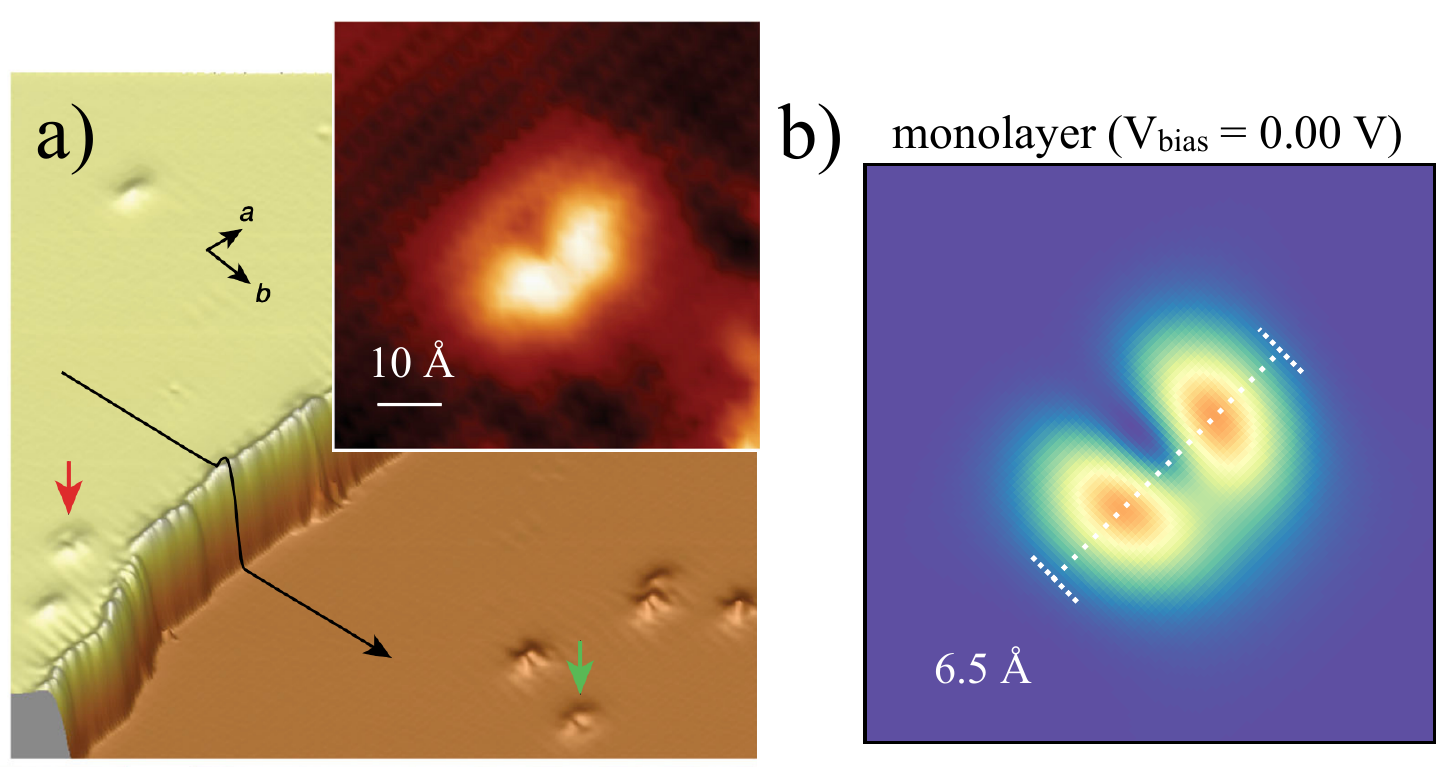}
\caption{
(a) Experimental STM image taken from~\citet{stm2}. The inset is a close-up of one of the defects.\\
(b) Simulated STM images of Te-adatoms on monolayer WTe$_2$ at zero-bias and a distance of 4.5 \AA \ from the surface. 
}\label{fig:four}
\end{figure}
\section{STM-images}
% The Te-adatoms have a low formation energy and are likely to be present in real monolayers of WTe$_2$, especially for crystals synthesized in Te-rich environments.~\cite{Tedefects,teflux}
Recent scanning tunneling microscopy (STM) experiments on WTe$_2$ have shown the existence of localized states on step edges in presence of un-identified defects on the surface of the material.~\cite{stm2}
Because of the layered nature of bulk WTe$_2$, a step edge can be modeled as a monolayer of WTe$_2$ on top of bulk WTe$_2$. 
As long as the coupling between the monolayer and the bulk is smaller than the band gap of the monolayer, the topological edge states of the monolayer are expected to be present on step edges, which is in agreement with the experimental findings. 
% Concordantly, the edge states observed in the experiment were identified as the topological edge states of a monolayer WTe$_2$.
The identification of the defects present in the STM images could yield important information about the nature of the step edge states as well as the stability of the TI state in monolayers of WTe$_2$ in general.
Based on the results of the previous section, we propose that the defects observed in these experiments are Te-adatoms, since their formation energy in a Te-rich environment is low, while they leave the topological structure of the material unperturbed.
To confirm this hypothesis, we have calculated STM images of a Te-adatom on a monolayer of WTe$_2$ at zero-bias [Fig.~\ref{fig:four}].
The experimental image shows the same characteristic shape as the calculated images,
i.e. a characteristic \mbox{"headphone"-like} shape of two maxima extended along the W-chain direction which resembles a distorted $p_x$ atomic orbital.~\cite{stm1,stm2}. 
\section{Conclusion}
We have studied the effect of two point defects on the topological properties of monolayers of WTe$_2$. We find that Te-adatoms have a low formation energy and that they do not affect the topological properties.
In contrast, Te-vacancies with a higher formation energy are likely to change the topological properties of the material since they strongly perturb the electronic structure close to the Fermi-level.
We predict that Te-adatom should be present in samples grown in Te-rich conditions, while Te-vacancies should have a consideratly lower concentration. 
Our findings are supported by recent STM images that measured topological edge states on step edges of WTe$_2$ in presence of un-identified defects, which we identify as Te-adatoms based on computed STM images.

\begin{acknowledgments}
We acknowledge fruitful discussions with B. Bradlyn, Z.J. Wang and C. Dreyer. RC was supported by DOE grants DE-SC0008626 and DE-SC0017865. 
This material is based upon work supported by the U.S. Department of Energy, Office of Science, Office of Advanced Scientific Computing Research and Office of Basic Energy Science, Scientific Discovery through Advanced Computing (SciDAC) program.
The Flatiron Institute is a division of the Simons Foundation.
\end{acknowledgments}
\appendix
\section{DFT calculations}
\label{app}
\begin{figure}[h]
\centering
\includegraphics[width=0.8\linewidth]{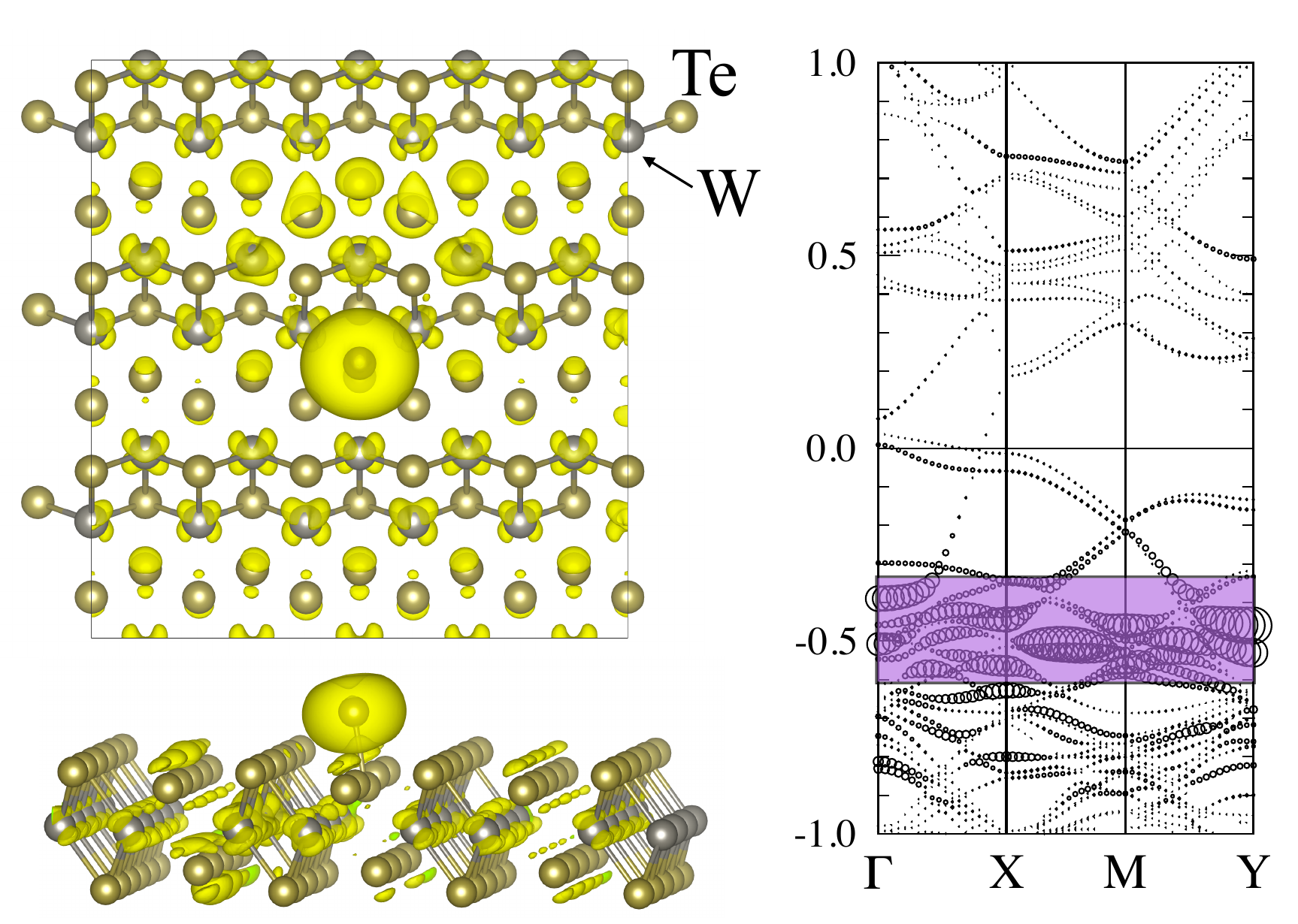}
\caption{Geometry and partial charge density of a relaxed $5\times3$ supercell of a Te-adatom on a monolayer of WTe$_2$. The energy window used to calculate the charge density is highlighted in the band structure of the supercell.
}\label{fig:app_ad}
\end{figure}
\begin{figure}[t]
\centering
\includegraphics[width=0.8\linewidth]{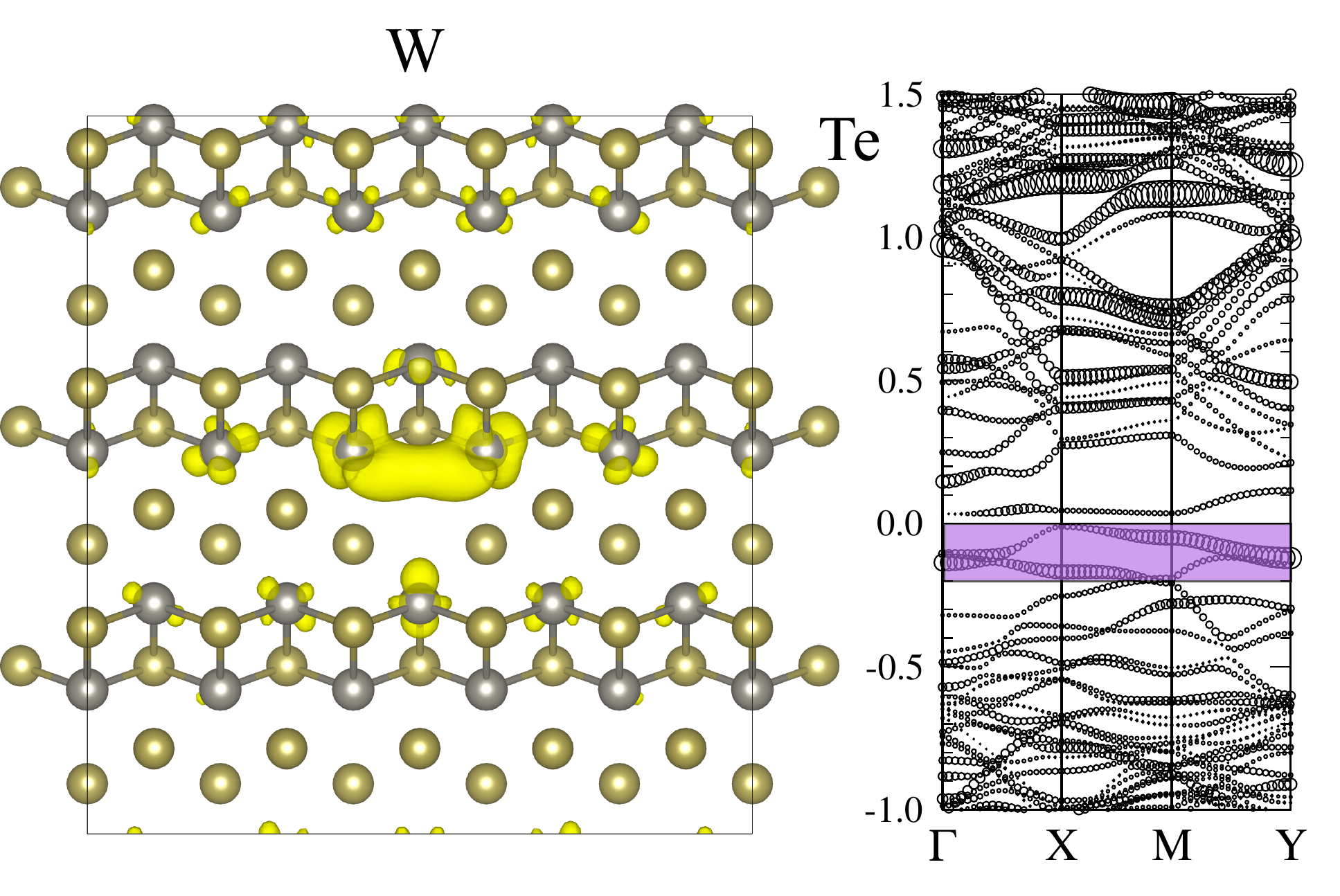}
\caption{Geometry and partial charge density of a relaxed $5\times3$ supercell of the Te-vacancy in a monolayer of WTe$_2$. The energy window used to calculate the charge density is highlighted in the band structure of the supercell.
}\label{fig:app_vac}
\end{figure}

We used the Kohn-Sham DFT-based electronic structure analysis tools implemented in the SIESTA (Spanish Initiative for Electronic Simulations with Thousands of Atoms) \cite{SIESTA} software package to study the atomic and electronic structure of WTe$_2$.
The generalized gradient approximation of Perdew, Burke, and Ernzerhof (GGA-PBE) \cite{GGA} for the exchange correlation functional was adopted. A double zeta plus polarization orbital basis set (DZP) was used to describe the valence electrons within the framework of a linear combination of numerical atomic orbitals (LCAO). All atomic coordinates were fully relaxed using the conjugate gradient (CG) algorithm until the energy and force convergence criteria of 10$^{-4}$ $eV$ and $0.02$ $eV/A$, respectively, were reached. All SIESTA calculations were performed on the Cori system available at the National Energy Research Scientific Computing (NERSC) centre. Due to the large number of atoms contained in the systems, the standard diagonalization (DIAGON) method in SIESTA based on the ScaLAPACK \cite{SCALAPACK} software package becomes prohibitively expensive. Therefore, we utilized the recently developed PEXSI (Pole Expansion and Selected Inversion) \cite{PEXI1,PEXI2,PEXI3,PEXI4} technique to reduce the computational time without sacrificing accuracy even for metallic systems.
We performed large supercell calculations for each type of defect described in the the main paper. Convergence of the total energy and defect formation energy was tested by varying the size of the supercells from $5\times3$ primitive cells up to $16\times9$.
Band structure and spectral function calculation were performed via the VASP package \cite{VASP,VASP2} based on the optimized geometries from the SIESTA calculations, using the GGA-PBE functional and a $4\times3\times1$ k-mesh with a plane-wave cut-off energy of 223 eV. 
Figure~\ref{fig:app_ad} shows the relaxed geometries of the Te-adatom obtained from the VASP calculations in a $5\times3$ supercell as well as the partial charge density from the bands with the strongest contribution of the adatom states.
Figure~\ref{fig:app_vac} shows the relaxed geometries of the Te-vacancy obtained from the VASP calculations in a $5\times3$ supercell as well as the partial charge density from the bands in a window of 0.2 eV below $E_F$.

\bibliography{wte2_defect.bib}

\end{document}